\documentclass[conference,a4paper]{IEEEtran}

\usepackage{ifpdf}

\usepackage{cite}

\usepackage{graphicx}

\usepackage{url}

\usepackage{color}
\usepackage{scrextend}
\usepackage{soul}
\usepackage[citebordercolor={1 1 1}, linkbordercolor={1 1 1}, urlbordercolor={1 1 1}]{hyperref}

\usepackage{setspace}

\usepackage{fancyhdr}
\definecolor{mygray}{gray}{0.65}
\renewcommand{\footrule}{\hbox to\headwidth{\leaders\hrule height \footrulewidth\hfill}}

\renewcommand{\footrulewidth}{0.75pt}

\DeclareRobustCommand*{\IEEEauthorrefmark}[1]{\raisebox{0pt}[0pt][0pt]{\textsuperscript{\footnotesize #1}}}

\begin{document}
\title{\textbf{On Large Ground Station Antennas as Potential Radar Targets for Biomass}}
\author{\IEEEauthorblockN{Cornelis G. M. van 't Klooster,\IEEEauthorrefmark{ \scriptsize(1)}   
Arnold van Ardenne\IEEEauthorrefmark{ \scriptsize(2)},   
}                                     
\\
\IEEEauthorblockA{
    \IEEEauthorrefmark{\scriptsize(1) }
     parttime Eindhoven University of Technology, Email: kvtklooster@gmail.com, C.G.v.t.Klooster@tue.nl}
    \IEEEauthorblockA{\IEEEauthorrefmark{\scriptsize(2) }
     parttime Astron, Oude Hoogeveensedijk 4, 7991 PD Dwingeloo, NL, Email: Ardenne@astron.nl}
}
%

\maketitle

\begin{abstract}
Radio-telescopes or ground-station antennas can, if pointed, act as a radar target with high radar cross-section (RCS). Space-based Synthetic Aperture Radar (SAR) data confirmed it at $5.3 GHz$ for a modified ground-station antenna. Operational ground-station antennas cannot be modified. The latter antennas might operate at frequencies well above the radar band. The radar signal could be scattered with high RCS from such an antenna, with less influence due to a load (receiver). The antenna geometry should be precisely known to derive its RCS. BIOMASS SAR operates near 435 MHz (P-band). Results are given, also for BIOMASS antenna itself. Large antennas in an array as in Westerbork are of potential interest, located on an East-West line, nearly perpendicular to ascending and descending polar orbits. Related material is discussed.
\end{abstract}
\section{Introduction}
\begin{figure}[!t]
\centering
\includegraphics[scale=0.99]{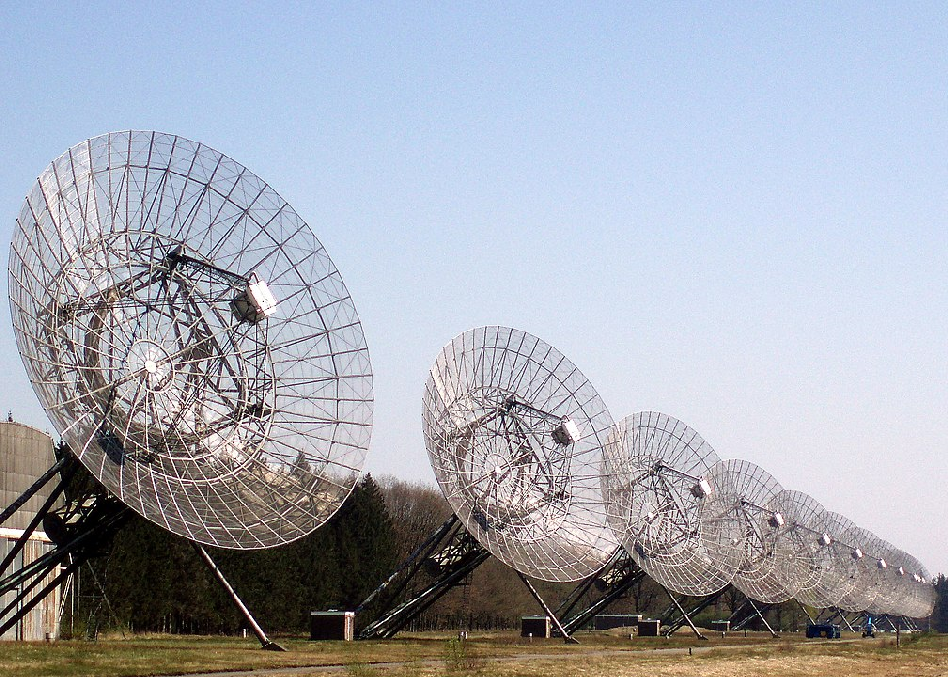}
\caption{Westerbork radio-telescope. (Copyright Astron \cite{Astron}).}
\label{fig_ESAworkshop KK1}
\end{figure}

A recent paper \cite{Quegan} discussed calibration aspects for BIOMASS. Its SAR will operate in a 6 MHz band near 435 MHz \cite{Biomass}. Radio-telescope antennas have been mentioned as radar target of opportunity with a requirement for antenna pointing but it has not been elaborated beyond one sentence in  \cite{Quegan}. This paper intends to provide additional information and suggests subjects for further investigation.
Ground station antennas have been mentioned \cite{Keen} as potential target for the SAR of the European Remote sensing Satellite (ERS) though with an uncertainty for the RCS value. There was a first mentioning of Westerbork radio telescope antennas (Fig. \ref{fig_ESAworkshop KK1}) in an internal memo in Estec Antenna Section in 1985. A potential additional one-way pattern measurement was mentioned for a part of ERS SAR elevation pattern, with different antennas in the East-West interferometer array. It has not been elaborated further at the time.\\
A development of active transponders has been undertaken for ERS \cite{Woode}. Dedicated antennas were needed with stringent side-lobe requirements derived at Estec to cope with direct and bi-static scattering effects nearby the transponder location. It led to a small, low side-lobe level antenna \cite{Medeiros} developed by ERA in UK.

\begin{figure}[!t]
\centering
\includegraphics[scale=0.9]{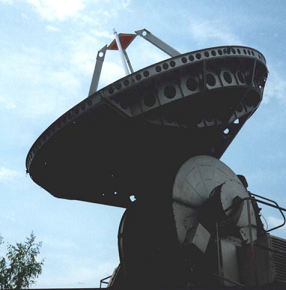}
\caption{Antenna with plate near Moscow, radar target used in \cite{Zakharov} with ERS SAR.}
\label{fig_ESAworkshop KK2}
\end{figure}

\begin{figure}[!t]
\centering
\includegraphics[scale=0.52]{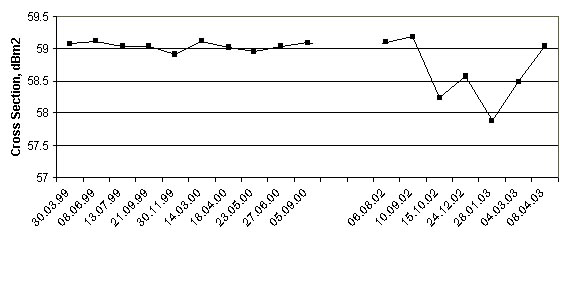}
\caption{RCS data for antenna Fig. \ref{fig_ESAworkshop KK2} \cite{Zakharov}.}.
\label{fig_ESAworkshop KK3}
\end{figure}

A very good absolute accuracy resulted for the SAR of ERS following a calibration campaign. The SAR has been used for assessment of RCS of ground-station antennas in Russia, at the time permitted by the ERS mission manager. First results have been presented in the CEOS working group meetings in Moscow \cite{CEOS1} and at Estec \cite{CEOS2}. The type of target was expected to be used also for a Russian SAR on the Priroda platform on MIR station. Further investigations confirmed RCS stability of a ground-station antenna observed with ERS. Interesting results were reported \cite{Zakharov} but for a modified antenna, see Fig. \ref{fig_ESAworkshop KK2}. The derived RCS stability is shown, see Fig. \ref{fig_ESAworkshop KK3} (from \cite{Zakharov}). A simple scattering structure is noticed in the focal area: just a small flat square plate. Such antenna modification is not allowed in an operating antenna. Such a plate assists to extend an angular domain for a high mono-static RCS value, but with limitations also. It just so happens that such a configuration scenario  compares with a focal plane scenario as present in a Westerbork antenna.

The use of ground-station antennas as a potential target came back again during the start of initial investigations of a potential P-band SAR mission. Suggestions have been put forward in \cite{Kloo1},  \cite{Kloo2}. The argument is that many operating ground-station antennas are "below cut-off" at P-band frequency. The antennas can offer potentially a large mono-static RCS with their scattering properties in lower bands, which are less modified by receiver reflection properties. Moreover, such a target benefits from antenna stability and pointing properties (designed for the shorter wavelength). SAR data have been collected at $5.3 GHz$ to explore the subject again \cite{Kloo1},  \cite{Kloo2} but really P-band data are needed. Initial estimations have been derived using Physical Optics (PO) \cite{Kloo2}. Results were also derived with the Method of Moment (MOM, \cite{Fasant1}, \cite{Fasant2}) but based on limited geometrical information. Nowadays various tools, based on MOM can handle scattering structures, which are very large in terms of wavelength. Predictions can be made using more detailed structural configuration aspects especially in the focal region. Accordingly a revisit is made here with a secondary objective to complement \cite{Quegan}.
\subsection{RCS aspects}
RCS is commonly defined as in Eq. \ref{eqKK1}.
\begin{equation}\label{eqKK1}
  \sigma = \lim \displaylimits_{r \to \infty} 4\pi r^2 \frac{\left|\vec{E}_s(\vec{r})\right|^2}{\left|\vec{E}_i(\vec{r})\right|^2}
\end{equation}
with position vector $\vec{r}$, the scattered field $\vec{E}_s(\vec{r})$ and the incident field $\vec{E}_i(\vec{r})$. A plane wave is assumed as the incoming field from a SAR satellite, properly normalised. Eq. \ref{eqKK1} presents a relation between incoming and outgoing power flux densities without  phase information.\\
The scattered field forms together with the incident field a total field $\vec{E}_{tot}$ (Eq. \ref{eqKK2}). The total field is complex and with its polarisation state and can be predicted in principle.
\begin{equation}\label{eqKK2}
  \vec{E}_{tot}(\vec{r}) =  \vec{E}_s(\vec{r}) + \vec{E}_i(\vec{r})
\end{equation}
The scattered field results from a surface current distribution induced on the scatterer by the incident plane wave field. The latter current distribution is determined by solving an electrical field integral equation (EFIE) derived on the basis of a boundary condition for the electric field. A solution is derived with the Method of Moments (MOM) and a scattered field can be derived subsequently. A physically large antenna or radio telescope is considered as a large structure. Precise configuration properties differ from antenna to antenna. Details need to be known, especially in the focal region. Accordingly one has to derive models for analysis of each antenna considered and antenna configurations can be quite different.\\
The complex scattered field with its polarisation state is available. It is potentially more desirable than just RCS values (no phase) to enable further detailed analyses. An antenna example is a 22 meter radio-telescope RT22 in \cite{Kloo2} but detailed configuration parameters were not available at the time. More detailed geometry data have been used here for RT22 \cite{Shulga}.\\
A 12 meter ALMA antenna and a 25 meter Westerbork antenna are considered with strut and blockage effects included. The Westerbork antenna is of particular interest, as there is a sort of cut-off aspect because of passive high-pass filters in all array elements in the focal plane array. The filter prevents signals below 800 MHz to enter front-end circuitry and thus to reduce interferences (Apertif focal plane array, 121 Vivaldi elements, \cite{Cappellen}, \cite{Astron}). Figure \ref{fig_ESAworkshop KK12} shows more details.\\
As a sort of curiosity the scattered field from the BIOMASS 12 meter antenna itself is shown (a P-band "selfie"). It is noted that the mono-static radar cross section is quite high with interesting polarisation properties.
\subsection{BIOMASS mission scenario}
BIOMASS SAR will fly at 666 kilometer altitude in a polar orbit, left-looking when ascending with an orbit-inclination of $98^\circ$. The synthetic aperture length is assumed here to be such that SAR processing is carried out over a synthetic aperture length ($\sim 24km$) corresponding to an angular width of about $1.8^\circ$ (well within the real aperture beam-width indeed). Assumed values serve to indicate principles, they are not necessarily the implemented values. The response from a radar target (for calibration or "of opportunity" ) should be stable and preferably not vary over the length of the synthetic aperture. A high RCS value is of interest well enough above the noise floor. A value of $85 dBm^2$ has been requested for calibration purposes but such a value is difficult to achieve with a large passive target like a radio telescope or Earth station antenna. Values achievable for passive targets can be available (more around $\sim 60 \; to \sim 65 dBm^2$ or less). Still such targets of opportunity can be of interest.
The resolution cell for the P-band SAR is close to $50$x$5m^2$ (elevation x azimuth), dictated by bandwidth and (circular) SAR antenna size as well as azimuthal $1^{st}$ ambiguity. A passive target like a $\simeq25m$ antenna will extend over more resolution cells. It is an aspect deserving more investigations.\\
An active radar transponder might realise higher RCS values, but it must then rely on high enough and stable amplification between (small) receive and transmit antenna with a low gain. "Small" in terms of wavelength presents still a physical size in P-band. An antenna has been proposed \cite{CEOS3} with potentially suitable side-lobe requirements to reduce multi-path and possibly bi-static scattering near the transponder. Antenna pattern aspects and dimensions are indicated in \cite{CEOS3}. A 2 meter aperture size is noted for a P-band horn as transponder antenna. The impact of bi-static scattering has not been reported. The horn proposed by DLR \cite{CEOS3} has also a low on-axis mono-static RCS and if it is well matched, it provides very little re-radiation. The horn is claimed to be patented but is in fact a well-known dual-mode corrugated horn. The active transponder provides an artificial target. Also it has finite dimensions and "fits in a resolution cell" of the BIOMASS SAR.

\section{Large antennas and estimations of the scattered field}
Every antenna acts as a scatterer, in fact scattering is needed in order to have reception and there is structural scattering. An antenna operating in higher frequency bands can act as a scatterer in P-band. We assume a perfect conducting antenna structure for the principles. Illuminating an antenna with a plane wave causes forward and backward scattering as well as re-radiation and in case of reception also absorption. If for instance the feed is below cut-off (or contains a filter) the antenna structure with feed still provides scattering, depending on precise antenna and feed structure but with less impact due to receiving circuitry. Below some estimations are given three types of radio telescope antennas and as a curiosity also for the Biomass antenna itself (a P-band "selfie"). The Westerbork antenna deserves further attention, as it can be an element in a group (array) and all elements have a high RCS value. It is very stable and an antenna is an array element of a larger array of such antennas at very precise distances on an East-West line, potentially an ideal scenario for targets of opportunity. It can provide a specific scenario with less elements, systematically positioned at certain distances East-West. The East-West line is close to the SAR elevation plane or $82^\circ$ or $98^\circ$ with respect to the BIOMASS inclined orbit plane.
\subsection{RT22 radio telescope}
The RT22 antenna has been observed by Envisat at $5.3GHz$ \cite{Kloo1}, \cite{Kloo2}. More precise RCS predictions are now given for P-band based on configuration details  from \cite{Shulga}. An assumption is made for dimensions of the focal container. A strut configuration and other focal area details are not included in the RT22 model. The resulting model for the Cassegrainian configuration is used for a comparison. The sub-reflector (diameter of $1.51m$) is towards a lower limit for Physical Optics (PO) in P-band. Three approaches have been used to serve a comparison, namely:
\begin{enumerate}
  \item PO for main- and sub-reflector,
  \item PO for main- and MOM for sub-reflector,
  \item MOM for the complete antenna configuration.
\end{enumerate}

\begin{figure}[!t]
\centering
\includegraphics[scale=0.8]{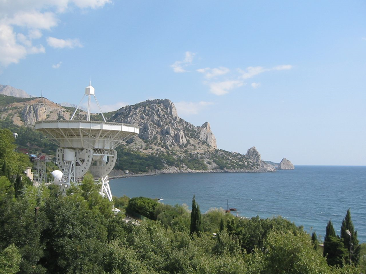}
\caption{Radio-telescope RT22 in Simeiz, Krim, \cite{Shulga}.}
\label{fig_ESAworkshop KK4}
\end{figure}

\begin{figure}[!t]
\centering
\includegraphics[scale=0.29]{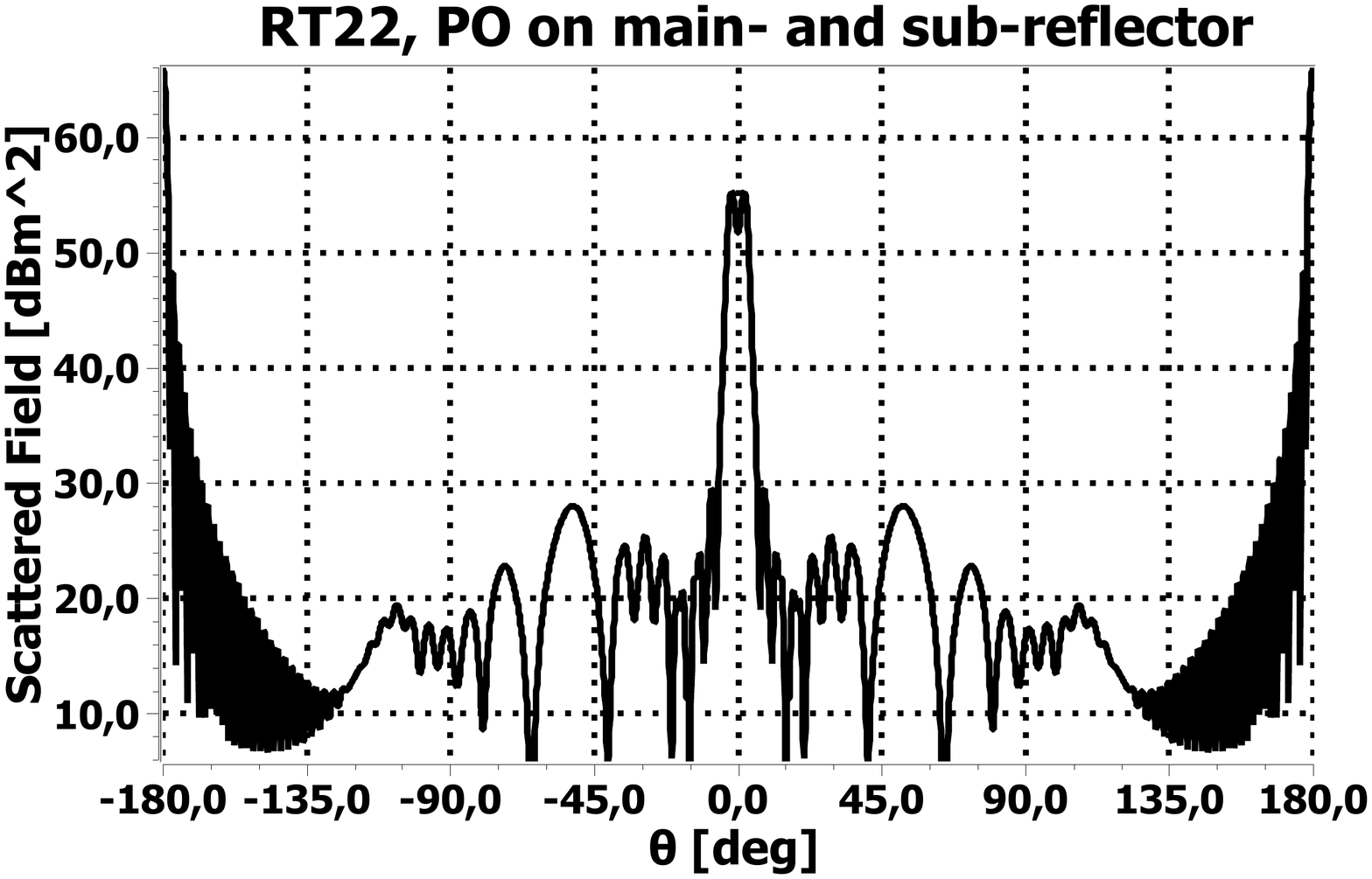}
\caption{RT22, PO on main- and sub-reflector.}
\label{fig_ESAworkshop KK5}
\end{figure}

\begin{figure}[!t]
\centering
\includegraphics[scale=0.29]{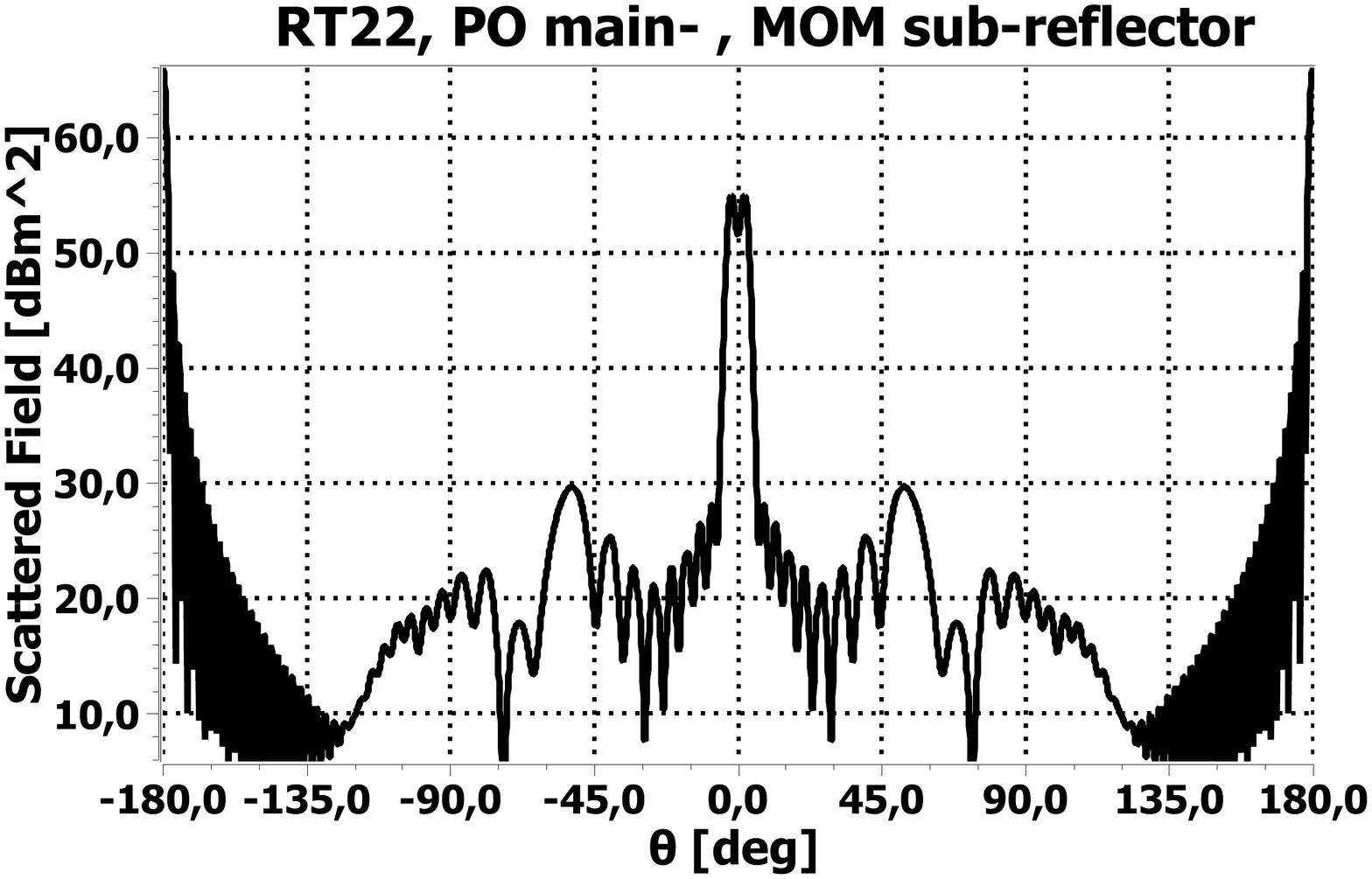}
\caption{RT22, main: PO, sub-reflector: MOM.}
\label{fig_ESAworkshop KK6}
\end{figure}

\begin{figure}[!t]
\centering
\includegraphics[scale=0.29]{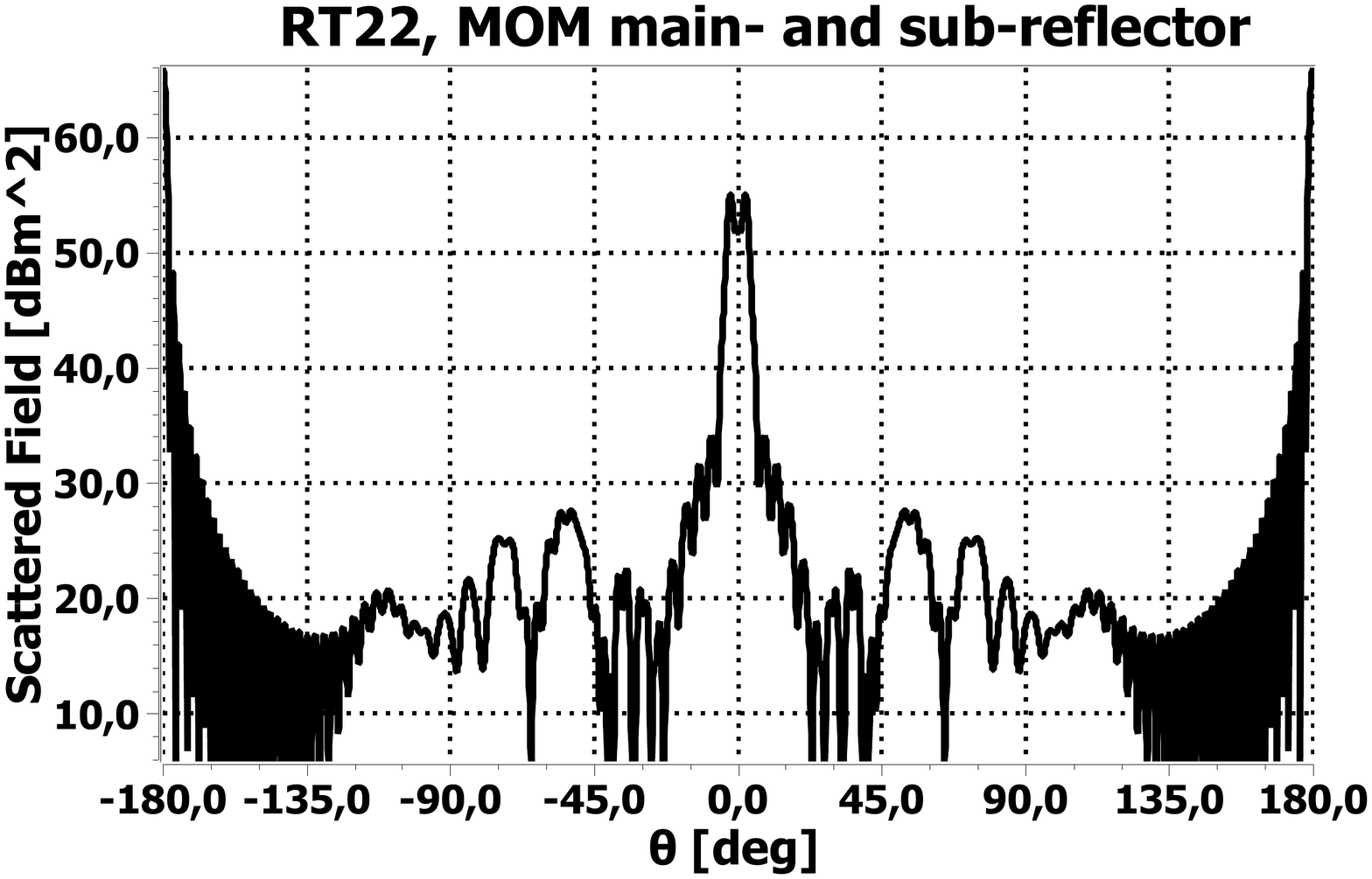}
\caption{RT22, MOM for complete antenna model.}
\label{fig_ESAworkshop KK7}
\end{figure}

Figure \ref{fig_ESAworkshop KK4} shows the RT22 radio-telescope. Figures \ref{fig_ESAworkshop KK5}, \ref{fig_ESAworkshop KK6} and \ref{fig_ESAworkshop KK7} present a result for the separate three cases, at $435MHz$. A comparable trend is observed with also differences. The incident plane wave is axially incident at $\theta = 0^\circ$, linearly polarised, Discrepancies become more pronounced for plane-wave incidence at increased angles $>\sim 1^\circ$ but not shown here. The forward scattering is comparable to the RCS of a disk of $22\;m$, as expected to assist to create a shadow zone behind the antenna. The PO analysis is still reasonable for a sub-reflector  $\sim 2 \lambda$ in size. A peculiar shape for the mono-static response is observed. A PO analysis at $5.3GHz$ shows a broadening effect at higher frequency as well, even more pronounced. One could recall measured results with Envisat (figure 11 in \cite{Kloo2}). Analyses with GRASP (well known antenna design tool, \cite{Ticra}) required verification of levels for calculation of the scattered field in terms of $dBm^2$. Classical targets with anticipated responses have been used (sphere, disk, corner reflector) to confirm such levels and allowed a reading directly in $dBm^2$. Results derived here supersede a MOM analysis result presented previously in \cite{Kloo1}, which was not correct. A supportive clarification by Ticra is acknowledged  \cite{Stig}.\\
The three approaches show mono-static on-axis responses several dB's lower than the forward scattering, with a specific shape and indicating a limitation to reach a high RCS for such antenna as RT22. Similar observations have been made for other antenna configurations (Yebes, FGAN, ESA 35m, Pico Veleta, IRAM) all with specific and different antenna properties. It is clear that detailed structural information is needed to obtain reliable predictions for targets of opportunity. One can predict responses for other (circularly, elliptically) polarised incoming plane waves.

Responses have been derived over a small angular interval, related to an available synthetic aperture length (a fictive assumption of 24 km). The mono-static complex response would be needed over the synthetic aperture length. It can also be expressed as a function of angle. The strategy sofar in earlier measurements has been to point a ground-station antenna towards the expected satellite position, when it is at the shortest range distance (predictable with orbit analysis). Pointing with accuracies of a few hundreds of a degree or less is usual daily routine for radio astronomers. A note can be made that in the SAR processing one has correction possibilities (for ERS-2 SAR at end-of-life it was needed as it could be a few degrees to cope with platform pointing problems). However if the precise complex scattered field is known and stable as function of angle, a SAR response would be derived, with a characteristic typical for the specific ground-station antenna considered, even when it is slightly larger than a few resolution cells in size.

\subsection{BIOMASS antenna}

\begin{figure}[!t]
\centering
\includegraphics[scale=0.29]{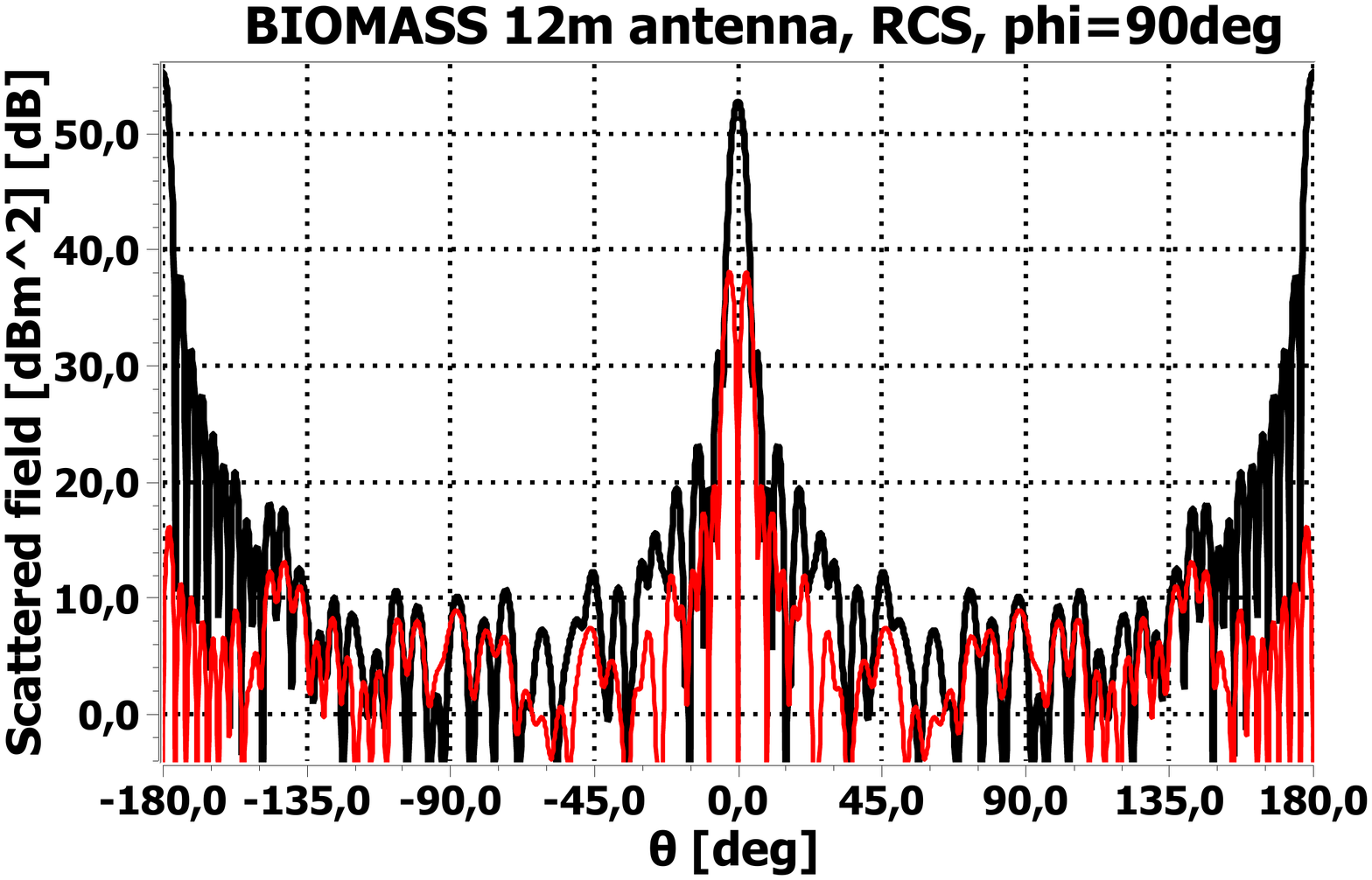}
\caption{BIOMASS antenna, P-band "selfie".}
\label{fig_ESAworkshop KK8}
\end{figure}

\begin{figure}[!t]
\centering
\includegraphics[scale=0.29]{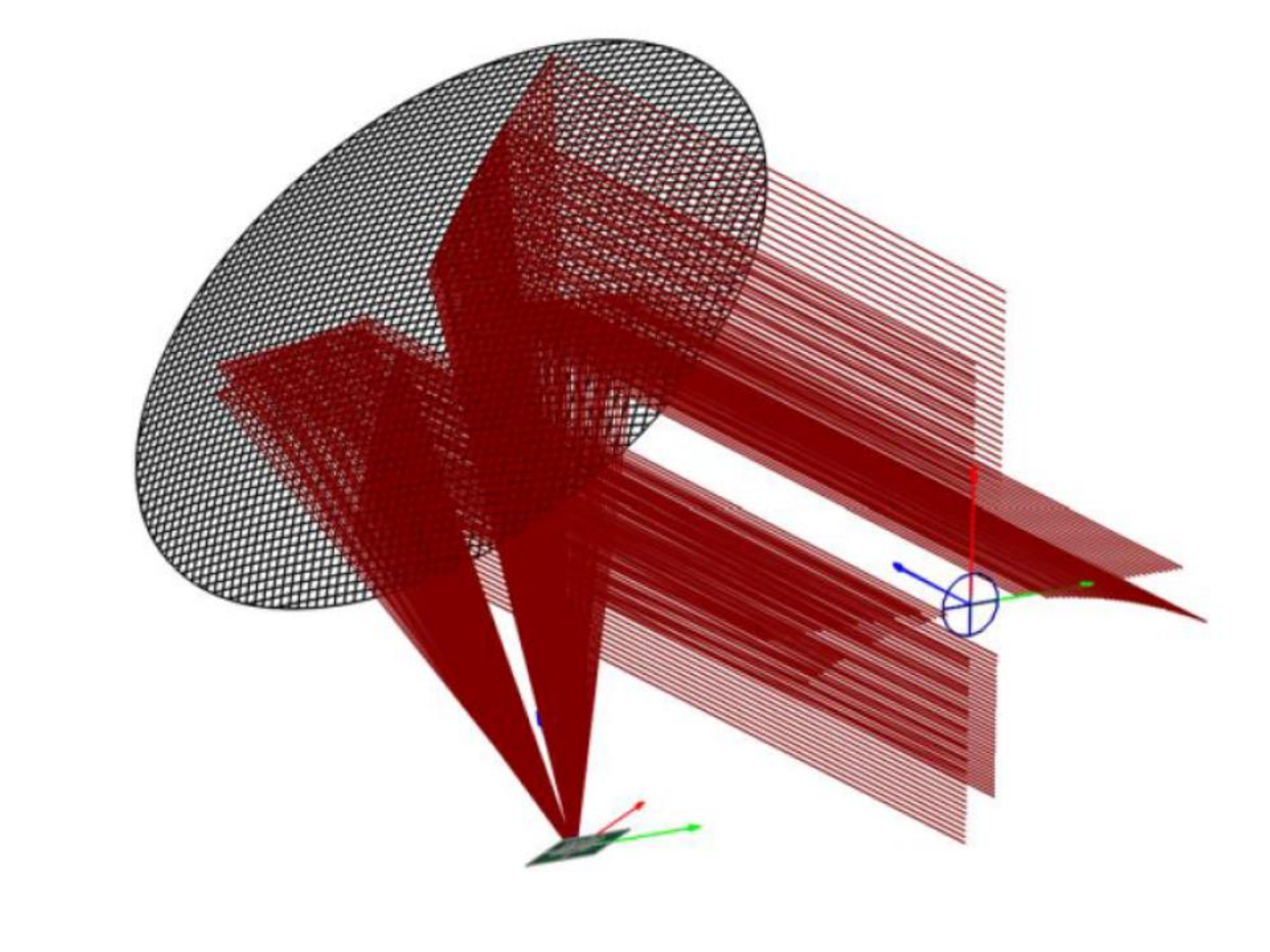}
\caption{BIOMASS in- and outgoing rays ($0^\circ$ and $90^\circ$ plane).}
\label{fig_ESAworkshop KK9}
\end{figure}

The BIOMASS offset antenna with its four patch element sub-array provides a scattered field as well to an incoming plane-wave. In the absence of detailed information an estimation is given, using a 12 meter offset reflector and a panel with four $0.48 \lambda $ patch elements, grounded in the centre on a front-plate. No feed network is considered (knowledge of the complete geometry will lead to the refined result). An incoming plane wave (linear polarisation assumed) is scattered and has when outgoing a cross polarisation which is nearly twice as high as one is used to for the (one-way) antenna pattern of such offset antenna. The mono-static RCS is rather high, with a clean response as there is no effect of struts and blockage (compared to other front-fed examples). Such antenna can act as a clean target of opportunity with high RCS, though - being in space - for a terrestrial based radar. It also has its specific polarisation response, due to the offset effect. The estimated scattered field for a plane perpendicular to the offset plane is shown as a "P-band \emph{selfie}" of BIOMASS antenna, see Fig. \ref{fig_ESAworkshop KK8}. The ray-tracing (see Fig. \ref{fig_ESAworkshop KK9}) shows the shift for the $0^\circ$ plane (offset plane) and the depolarisation in the $90^\circ$ plane, thus a very interesting polarisation sensitive target of opportunity but it would be there in space (so suitable for a radar on the ground...).
\subsection{ALMA radio telescope antenna}
\begin{figure}[!t]
\centering
\includegraphics[scale=0.29]{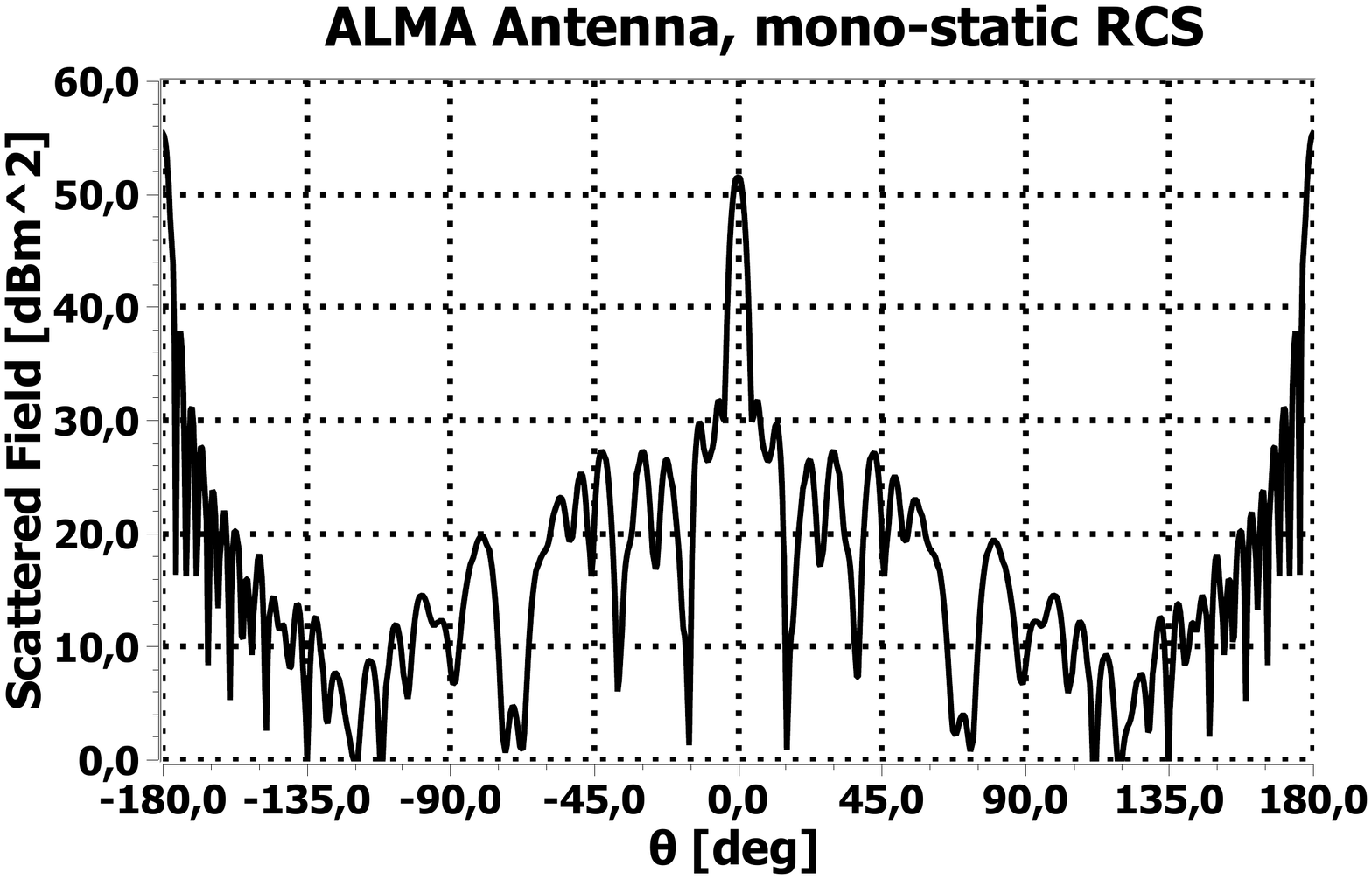}
\caption{ALMA radio-telescope, mono-static response (435 MHz).}
\label{fig_ESAworkshop KK10}
\end{figure}

\begin{figure}[!t]
\centering
\includegraphics[scale=0.29]{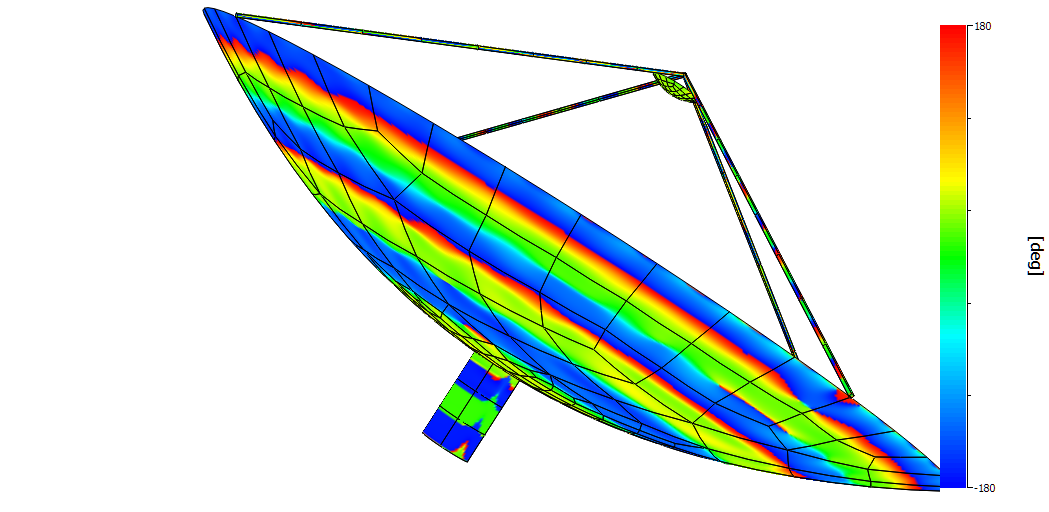}
\caption{ALMA Copol Current component, relative phase.}
\label{fig_ESAworkshop KK9alma}
\end{figure}

An ALMA antenna has been modelled with struts included and a circular cavity between secondary focus and main reflector. Geometry parameters can be found in the ALMA memo-series (ESO, NRAO) and experience assists here further. The "European" antenna has been assumed with single struts to the edge of the main reflector. The sub-reflector and cavity diameter are near to $\sim 1 \lambda$. Accordingly its scattered field cannot be calculated accurately with PO and MOM must be used. A high mono-static RCS results for a complete antenna as target. A result for the scattered field is shown in Fig. \ref{fig_ESAworkshop KK10} for a mono-static response and axial incident linearly polarised plane wave). The on axis RCS value is near to $51.5 dBm^2$ and includes effects of struts and blockage. Just an x-directed current component (phase) is shown in Fig. \ref{fig_ESAworkshop KK9alma}. An additional analysis confirmed a stable level over an angular range of $1.8^\circ$ corresponding to earlier indicated assumptions. Such a test showed stability or relative constant level in back-scattering. The forward scattering corresponds closely to a value of $55.3dBm^2$ as for a circular disk of $12m$ with a mono-static back scattering $\sim 3.8dB$ below that level.\\
It is interesting to compare differences between an ALMA 12 meter antenna and a 12 m offset antenna as for BIOMASS (see Fig. \ref{fig_ESAworkshop KK8}). Differences in level and polarisation properties are noticed: antennas with similar diameter have quite different scattering properties.

\subsection{Westerbork antenna}
\begin{figure}[!t]
\centering
\includegraphics[scale=0.55]{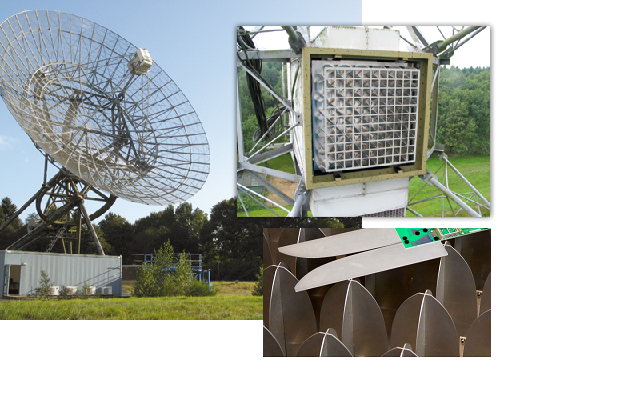}
\caption{Westerbork antenna, focal array, element and filter (Copyright Astron)}
\label{fig_ESAworkshop KK11}
\end{figure}

\begin{figure}[!t]
\centering
\includegraphics[scale=0.27]{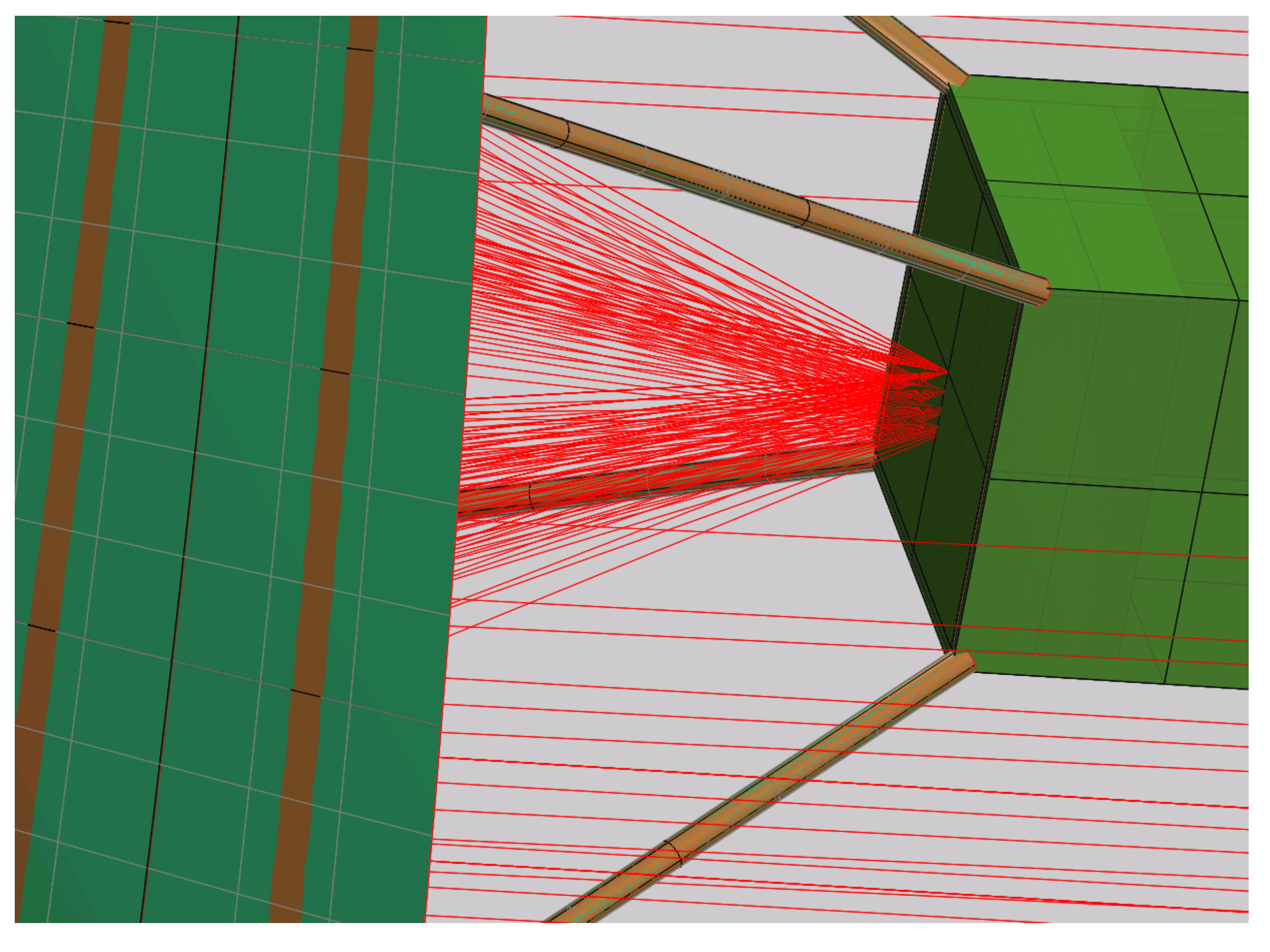}
\caption{Model focal container with indication with rays}
\label{fig_ESAworkshop KK12}
\end{figure}

\begin{figure}[!t]
\centering
\includegraphics[scale=0.30]{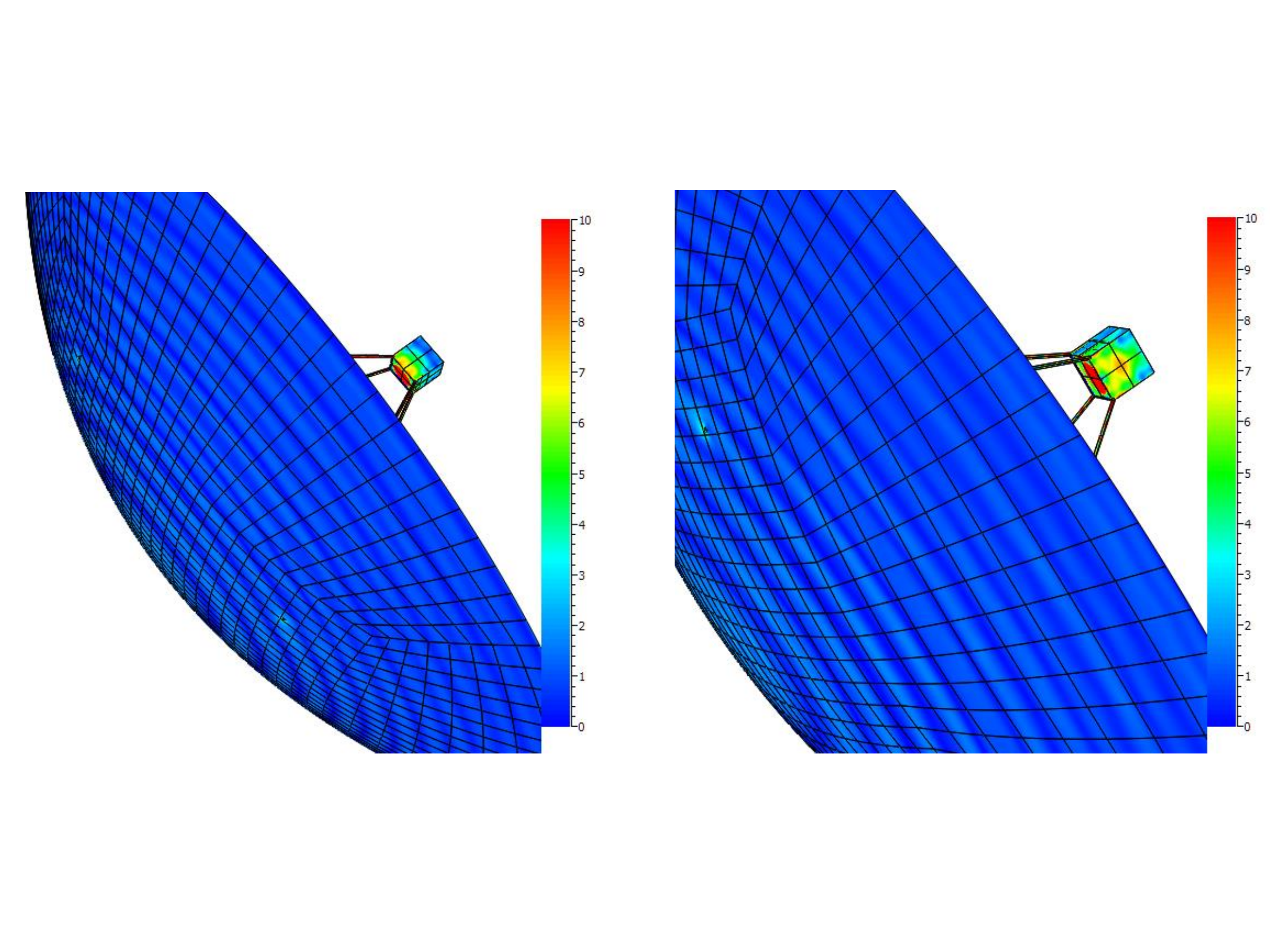}
\caption{Currents for $0^\circ$ (left) and $0.9^\circ$ incidence}
\label{fig_ESAworkshop KK13}
\end{figure}

\begin{figure}[!t]
\centering
\includegraphics[scale=0.30]{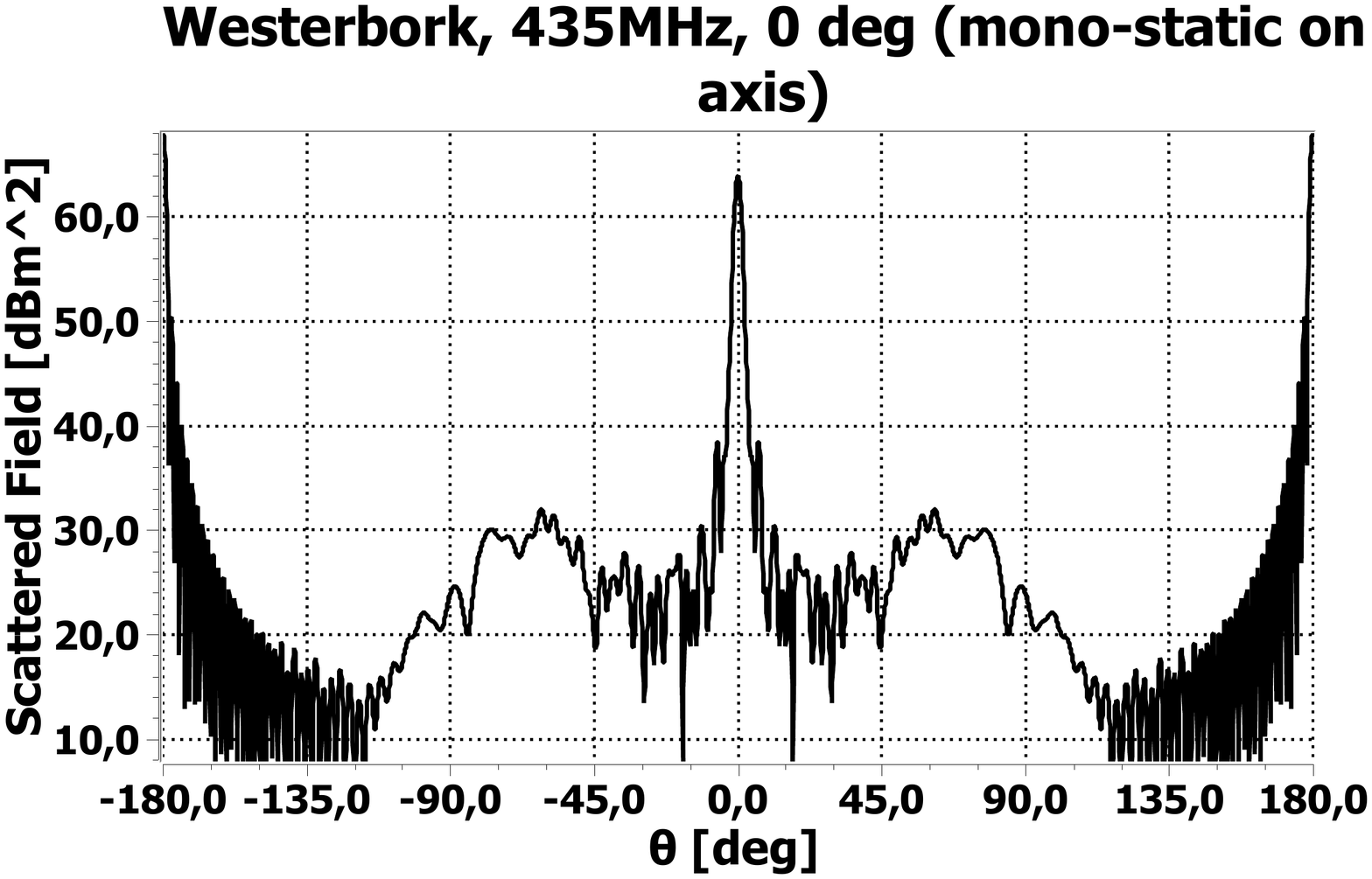}
\caption{Westerbork, scattered field, $0^\circ$ incidence}
\label{fig_ESAworkshop KK14}
\end{figure}

\begin{figure}[!t]
\centering
\includegraphics[scale=0.30]{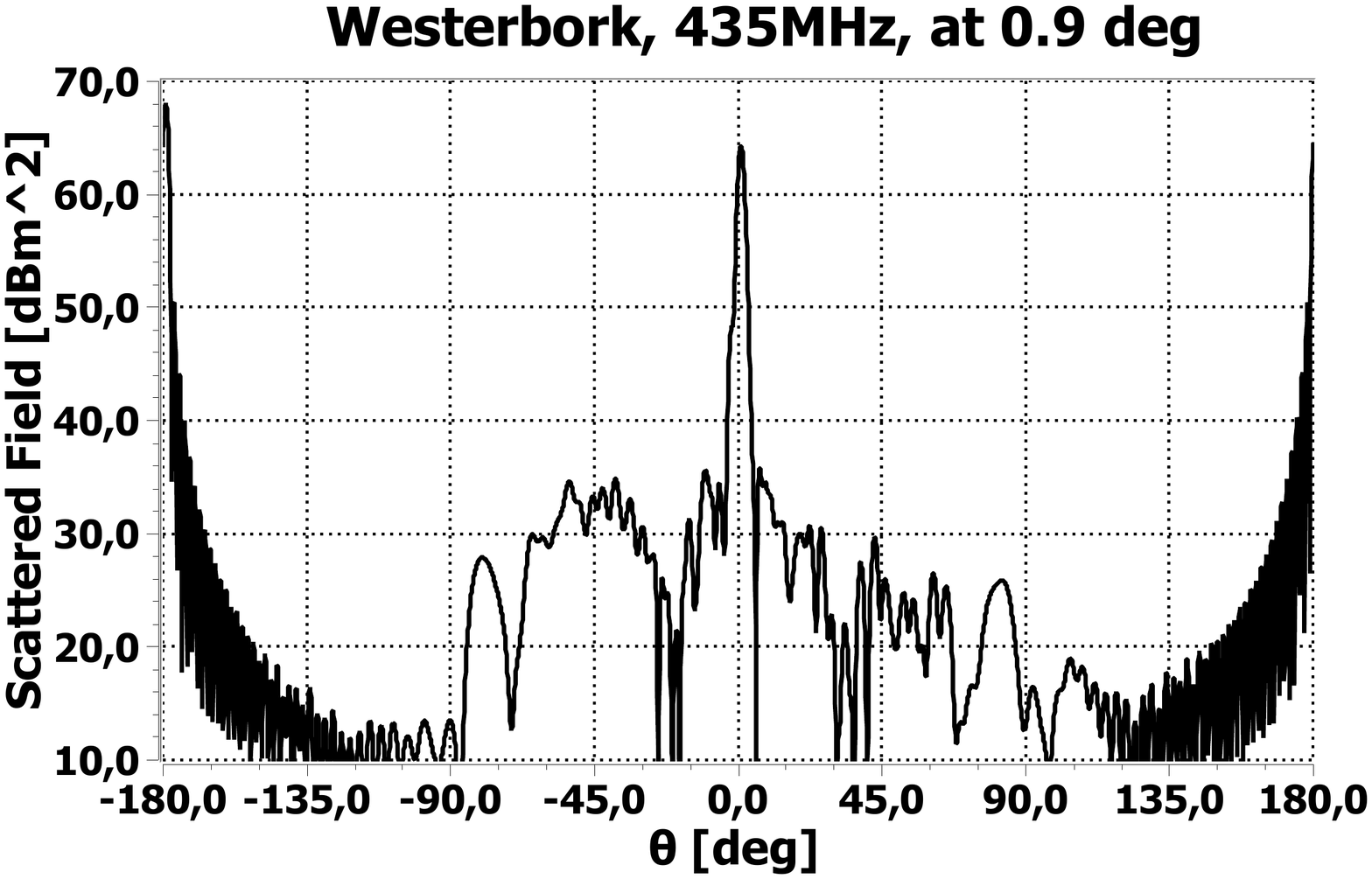}
\caption{Westerbork, scattered field, $0.9^\circ$ incidence}
\label{fig_ESAworkshop KK15}
\end{figure}

\begin{figure}[!t]
\centering
\includegraphics[scale=0.28]{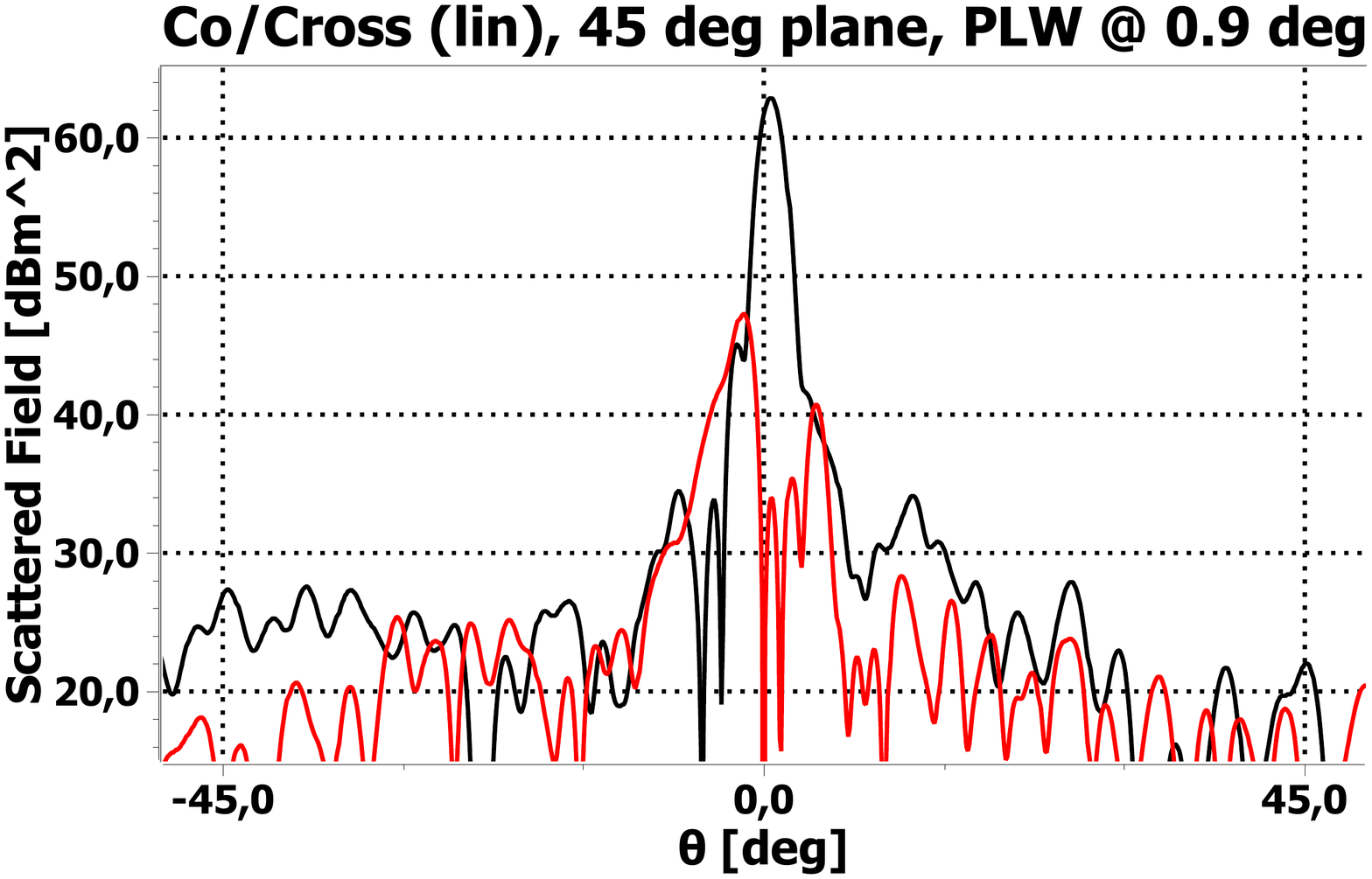}
\caption{Westerbork, scattered field, $45^\circ$-plane, $0.9^\circ$ incidence}
\label{fig_ESAworkshop KK16}
\end{figure}

A model was made for a Westerbork antenna, based on geometry parameters available from various sources, mainly from \cite{Astron}. The focal plane array is specific and contains 121 Vivaldi elements with an 800 MHz high-pass filter in each element \cite{Cappellen}. The filter prevents 435 MHz to enter the front-end. An assumption is made that incoming waves (plane wave spectrum for the angle subtended by the reflector) are reflected. The precise and effective impact of the focal array with filters needs to be analysed, this time for structural scattering and reradiation at P-band. Tools have been developed potentially useful \cite{Astron}, \cite{Maaskant} at Astron and Eindhoven University of Technology. An assumption here is taken that there is a perfect reflector positioned at (assumption) the focal distance, which is not true but serves the purpose. Localisation with respect to the focus has impact on the scattered field and precise analysis is necessary. The reality is probably a location of a quasi-planar equivalent surface slightly in front of the focus towards the reflector.\\
Array elements are spaced in a diagonal grid. Element spacings are near to $8 cm$ in the $45^\circ$ direction, providing a quasi wire-grid loading for the incident plane wave spectrum. Quasi because of the filter located nearly $\frac{1}{4}\lambda$ away from the front surface, another reason to analyse in more detail. As mentioned it is simulated with a reflecting plate in the focus.
Figure \ref{fig_ESAworkshop KK11} illustrates the focal plane array and its elements. Note the structural support, which is not included in full detail in the model either and which will have its effects at $435MHz$, more than for the assumed model here.
The effect of scattering by struts leads to a reduction for the on-axis mono-static RCS ($ \sim 0.5dB$ was noted). A variation in level of the scattered field has been noticed for $\theta > \sim 1^\circ$. It is also a reason why pointing should be accurate.\\
Figure \ref{fig_ESAworkshop KK12} shows the sketch for rays corresponding to plane wave incidence and $0.3^\circ, \; 0.6^\circ, \; 0.9^\circ$ incidence. It is observed, how expected smearing starts for off-axis directions. The maximum and first side-lobe (partly) of a focal field (nearly an Airy function) are noticed on top of the focal plane which simulates the focal array front-face. It becomes clearly more asymmetric when going off-axis with the incident plane wave direction. Nevertheless stability of the RCS level has been verified over an $1.8^\circ$ angular interval with the antenna model. A representation of the current distribution shows such effects in detail when going off-axis (see Fig. \ref{fig_ESAworkshop KK13}), with on-axis plane-wave incidence on the left and incidence at $0.9^\circ$ on the right. One observes the deformation of intensity of currents on the focal plate (lateral shift) as well as the appearing change in intensity on the sidewall of the focal container. The scattered field starts to increase slightly when going to larger incidence angles before dropping down.\\
The scattered field is predicted for $0^\circ$ and $0.9^\circ$ incidence (see Fig. \ref{fig_ESAworkshop KK14} and Fig. \ref{fig_ESAworkshop KK15} respectively), both at $435 MHz$. Figure \ref{fig_ESAworkshop KK16} shows the response for a $45^\circ$ plane for a plane wave incident at $0.9^\circ$ in the $0^\circ$ plane. A polarisation ratio is noted near -30dB in the direction back to the satellite.
It shall be investigated for the range of directions as occurring for the synthetic aperture length and beyond (related to processing windows chosen as well in SAR processing. The satellite moves as a sensor along the orbit and samples in a time-sequential manner the synthetic aperture).
\subsection{The Westerbork East-West array configuration}
\begin{figure}[!t]
\centering
\includegraphics[scale=0.32]{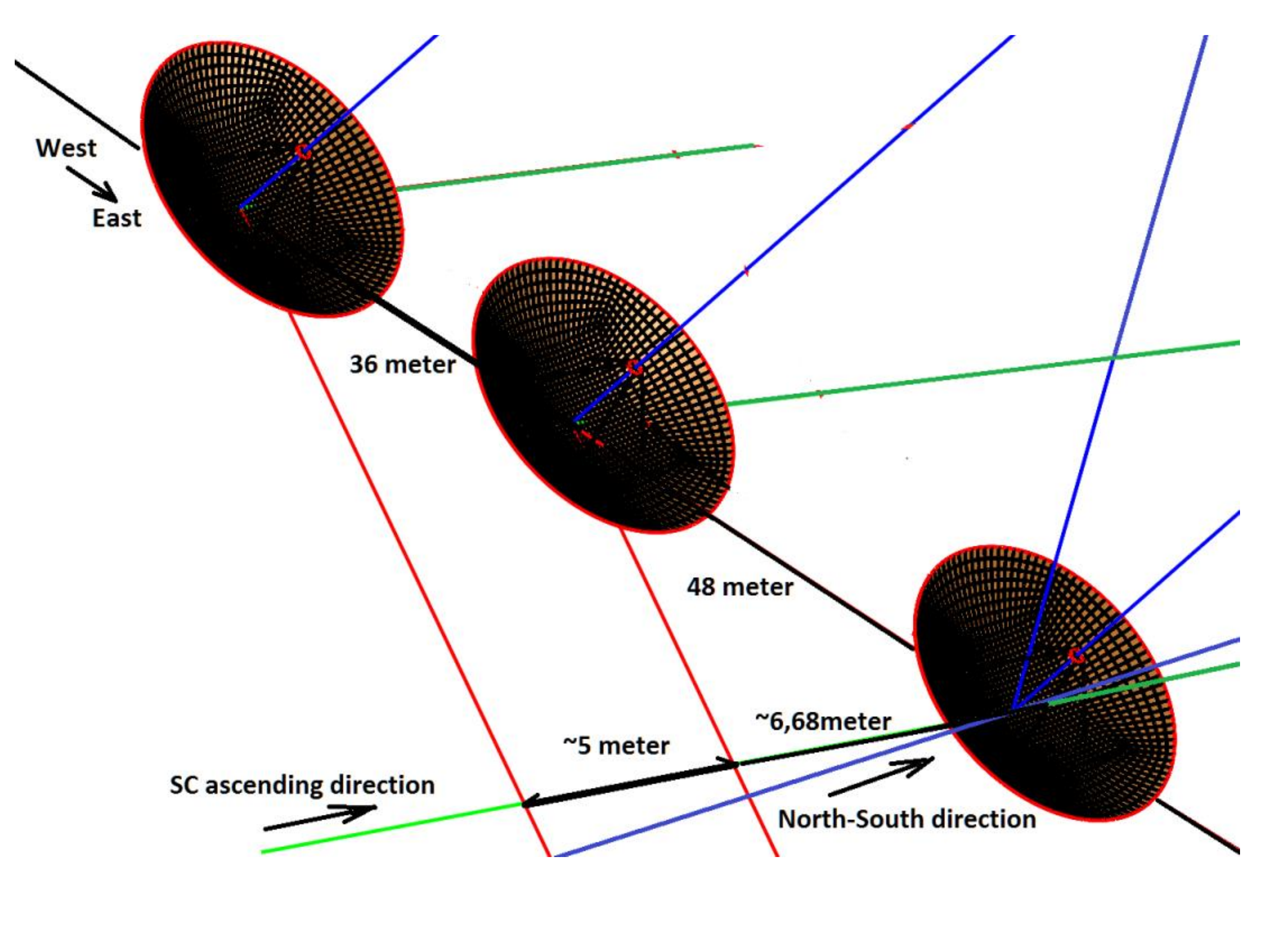}
\caption{Three antennas spaced 36m and 48 m respectively}
\label{fig_ESAworkshop KK17}
\end{figure}

A systematically located and specific target scenario is of potential interest. The Westerbork array of 14 radio telescope antennas (or a sub-set) could act as such a specific target scenario with specific and accurate inter-element distances, extending to $2.7 km$ \cite{Astron}.\\ Most antennas are spaced at 144 m spacing, the interval between antenna 10 and a next antenna can be as small as 36 meter, it can be followed by a possible spacing of 48 meter. The remaining two antennas (13 and 14) are at a larger distances. A spacing of 36 meter is formally below a resolving capability of the BIOMASS SAR (elevation direction). The distance projected on the ground azimuthal track is close to an azimuth resolution capability also. The East-West line is very close to the SAR elevation plane. A SAR response for two antennas spaced 36 meter becomes a complex addition, which could be estimated (complex, amplitude and phase and polarisation-state). A maximum would appear for in-phase addition, but it exactly depends on orbit configuration and SAR processing. BIOMASS SAR would not be able to resolve such targets completely. It would make the target scenario unique. One can select more targets obviously, like the interval of 48 meter distance and also other distances, but then at 144 meter.\\
Accordingly a systematic specific target scenario ("of opportunity") results. Each Westerbork antenna is mounted on an alt-azimuth positioner. All antennas are nicely aligned on polar axes for East-West movement. It gives similarity in effects on the East-West line as caused by structure and struts for each antenna with symmetries potentially to be exploited for ascending and descending orbits (it provided perhaps the most accurate polarimeter already in the late seventies of last century). Figure \ref{fig_ESAworkshop KK17} sketches three 25 meter Westerbork antennas relatively spaced (36 m and 48 m) as targets of opportunity on an East-West line and the assumed direction of a satellite track on ground. Obviously the side-looking SAR satellite passes to the East or to the West depending its (ascending or descending) orbit.
\subsection{An RFI aspect}
A demand exist not to use P-band SAR within coverage areas of certain terrestrial radars, related to ITU frequency allocations. Westerbork is geographically potentially within such a coverage area. It would prevent transmissions by BIOMASS over the Westerbork location.
With a satellite-speed in orbit near to $7.2 \frac{km}{sec}$ only a transmission of some tens of seconds would be needed over Westerbork. If so possible a collocation with the active and or DLR transponder could be another asset, perhaps even on site,

\section{Concluding remarks}
Scattering properties of large radio telescope antennas have been discussed and new analysis results have been presented for 3 radio telescope antennas and for the BIOMASS SAR antenna itself. As such it is considered to complement material presented in \cite{Quegan} where only one sentence has been devoted to large ground station antennas as target of opportunity with pointing aspects. It has been found that there can be large RCS values associated, but there can be limits for maximum levels as well as peculiar shapes for mono-static responses, which can be on the other hand of interest as well. It depends on precise antenna structural configuration information: detailed geometrical information is needed for prediction tools.\\
High dynamic ranges occur in analyses and specific small detailed information can be needed in focal regions, thus reflectors large in terms of wavelength next to details small in terms of wavelength.\\
Obviously convergence aspects need to be checked. Using different methods like PO and MOM or combinations assists to obtain further confidence in predictions. Such assessment can be useful again every time when another antenna configuration is investigated.\\
One would need to obtain transmission time over Westerbork because of a specific frequency allocation situation. But the Westerbork radio-telescope array remains an interesting target scenario (of opportunity) with antennas at precise specific distances, with interesting resolution aspects related and with antenna dimensions larger and smaller than resolution cell dimensions of the BIOMASS-SAR.\\
In conclusion it might be of interest to investigate this and various related aspects as well in more detail.

\section{Acknowledgement}
A special acknowledgement goes to Eindhoven University of Technology, Prof.Dr.Ir. A.B. Smolders in particular, to enable development and derivation of results presented here. Ir. Björn Rommen of ESA Estec is acknowledged for discussions on (BIOMASS-)SAR.\\

\section{Bibliography}

\begin{spacing}{1.2}

\end{spacing}


\begin{thebibliography}{1}

\bibitem{Quegan}
S. Quegan, M. Lomas, K. P. Papathanassiou, J-S. Kim, S. Tebaldini, D. Giudici, M. Scagliola, P. Guccione, J. Dall, P. Dubois-Fernandez and P. Paillou. Calibration challenges for the BIOMASS P-Band SAR instrument, \emph{IEEE IGARSS Conf.}, Valencia, Spain, 22-27 July, 2018.

\bibitem{Biomass}
Report BIOMASS, ESA \url{http://www.esa.int/Our_Activities/Observing_the_Earth/ESA_s_Biomass_satellite_goes_ahead}.

\bibitem{Keen}
K.M. Keen. Use of radio telescopes or satellite earth station antennas as ultra-high scattering
cross-section calibration targets for spaceborne remote sensing radars. \emph{Electr. Lett.}, vol. 19, no. 6, pp. 225-226, 17 March, 1983.

\bibitem{Woode}
H.D. Jackson, A. Woode. Development of the ERS-1 active radar calibration unit. \emph{IEEE Trans. Microw. Theory and Techn.},  vol. 40, no. 6, pp. 1063-632, June 1992.

\bibitem{Medeiros}
F. Medeiros, M. Simmons, K. van 't Klooster, M. Chadwick. Small low-sidelobe reflector antenna for the transponder for ERS-1. \emph{IEEE APS}, Dallas, 7-11 May, 1990.

\bibitem{Zakharov}
A.I. Zakharov, P.A. Zherdev, M. Borisov, A. Sokolov, C.G.M. van 't Klooster. On the Stability of Large Antennas as Calibration Targets", Proc. IGARSS Conf. 2003.

\bibitem{CEOS1}
\url{http://sarcv.ceos.org/documents/}

\bibitem{CEOS2}
Van’t Klooster C.G.M., C.H. Buck., P.A. Jerdev, M.M. Borisov, V.I.Gusevsky and A.I.Zakharov. “On the use of
ground-based parabolic reflector antennas for external calibration of spaceborne SAR”, \emph{CEOS SAR
Workshop}, Noordwijk NL, ESA WPP-138, pp. 237-240, 1998.

\bibitem{CEOS3}
Jens Reimann, Marco Schwerdt, Sravan Kumar Aitha and Manfred Zink.
\url{http://sarcv.ceos.org/site_media/media/documents/S22.1_Reimann.pdf}

\bibitem{Ticra}
\url{http://www.ticra.com}

\bibitem{Fasant1}
\url{http://www.fasant.com/}

\bibitem{Fasant2}
F. Catedra, communication, 2012.

\bibitem{Stig}
Stig Soerensen, Ticra, communication, 2018.

\bibitem{Kloo1}
C.G.M. van 't Klooster. About ground-station antennas as ‘radar target’ for P-Band synthetic aperture radars in space. \emph{IEEE APSURSI}, Spokane, 3-8 July 2011.

\bibitem{Kloo2}
Kees van't Klooster, Björn Rommen, Felipe Cátedra. Ground-based reflector antennas observed with space-based Synthetic Aperture Radar. \emph{6th European EUCAP Conf}. Prague, pp 1086-1090, 2012.


bibitem{Ticra}
\url{http:\\www.ticra.com}

\bibitem{Shulga}
A.V. Antiyufejev, S.Yu, Zubrin, V.V. Mishenko, I.I. Zinshenko, A.E Volvach, V.M. Shulga. Investigation of the antenna parameters of RT22 KrAO at a wavelength of 3.42mm. \emph{RadioFysika and RadioAstronomia}, 2009, T.14, No4, c.345-352 (in Russian).

\bibitem{Cappellen}
W.A. van Cappellen, J.G. Bij de Vaate. Status update on APERTIF, Phased array feeds for the Westerbork radio-telescope. \emph{URSI General Assembly}. Beijing, China, 2014.


\bibitem{Maaskant}
R. Maaskant. Analysis of Large Antenna Systems, \emph{PhD Thesis TU Eindhoven}. Eindhoven, NL, 2010.

\bibitem{Astron}
\url{http://www.astron.nl}


\end{thebibliography}
\end{document}